\documentstyle[12pt]{article}
\def\trg{{\triangleright}}
\def\ker{\mathop{\rm ker}}
\def\rtimes{{\mbox{$\times\!\rule{0.3pt}{1.1ex}\,$}}}
\def\A{{\cal A}}
\def\G{{\Gamma (\A)}}
\def\O{{\Omega (\A)}}
\def\R{{\cal R}}
\def\N{{\cal N}_{\R}}
\def\U{{\cal U}}
\def\T{{\cal T}_{\R}}
\def\K{{\cal K}}
\def\KR{\K / \R}
\def\GN{\Gamma _{\R}}
\def\ON{\Omega_{\R}}
\def\IA{{1_{\A}}}
\def\IU{{1_{\U}}}
\def\DA{{\Delta_{\A}}}

\def\I{{\mbox{\boldmath $i$}}}
\def\lie{\mbox{\pounds}}
\def\inprod#1#2{\left\langle #1, #2\right\rangle}
\def\dinprod#1#2{\left\langle \inprod{#1}{#2} \right\rangle}
\def\comm#1#2{\left[ #1, #2\right]}
\def\acomm#1#2{\left\{ #1, #2\right\}}
\def\id{\mathop{\rm id}}
\def\bid{\mbox{\bf id}}
\def\dg{{\mbox{\boldmath $\delta$}}}
\def\om{{\omega}}
\def\al{{\alpha}}
\def\ad{{\stackrel{\mbox{\scriptsize ad}}{\triangleright}}}
\def\tad{{\stackrel{\mbox{\tiny ad}}{\triangleright}}}
\def\Ad{{\Delta^{\mbox{\scriptsize Ad}}}}
\def\ie{\mbox{{\it i.e.} }}
\def\eg{\mbox{{\it e.g.} }}
\def\matrix#1#2#3{#1 ^{#2}{}_{#3}}
\begin{document}
\begin{titlepage}
\begin{center}
February 23, 1994	\hfill    LBL-33655\\
			\hfill    UCB-PTH-93/36\\

\vskip .5in

{\large \bf Universal and Generalized Cartan Calculus on Hopf
Algebras}\footnote{This work was supported in part by the Director,
Office of Energy Research, Office of High Energy and Nuclear Physics,
Division of High Energy Physics of the U.S. Department of Energy under
Contract DE-AC03-76SF00098 and in part by the National Science
Foundation under grants PHY-90-21139 and PHY-89-04035.}

\vskip .5in

Peter Schupp\footnote{schupp@physics.berkeley.edu} and Paul
Watts\footnote{watts@theorm.lbl.gov}

\vskip .5in

{\em Department of Physics\\
University of California\\
and\\
Theoretical Physics Group\\
Lawrence Berkeley Laboratory\\
University of California\\
Berkeley, California 94720}
\end{center}

\vskip .5in

\begin{abstract}

We extend the universal differential calculus on an arbitrary Hopf
algebra to a ``universal Cartan calculus''.  This is accomplished by
introducing inner derivations and Lie derivatives which act on the
elements of the universal differential envelope.  A new algebra is
formulated by incorporating these new objects into the universal
differential calculus together with consistent commutation relations.
We also explain how to include nontrivial commutation relations into
this formulation to obtain the ``generalized Cartan calculus''.

\end{abstract}

\end{titlepage}

\newpage
\renewcommand{\thepage}{\arabic{page}}
\setcounter{page}{1}

\section{Introduction}

The question of how to endow a quantum group with a differential
geometry has been studied extensively \cite{Wor}-\cite{CW}.  Most of
these approaches, however, are rather specific: many papers dealing
with the subject consider the quantum group in question as defined by
its R-matrix, and others limit themselves to particular cases.  In
this paper, we attempt a more abstract and general formulation, where
we consider an {\em arbitrary} Hopf algebra rather than a quantum
group as the basis for our differential geometry.

The approach we take starts with a Hopf algebra $\A$ and its
associated universal differential calculus $(\O ,\dg )$
\cite{Connes}.  By introducing another Hopf algebra $\U$ which is
dually paired with $\A$, we then construct a larger class of
generalized derivations (given in terms of elements of $\U$) which act
on $\O$; these play the roles of Lie derivatives and inner
derivations.  It is then possible to extend $\O$ to a larger algebra
by finding the commutation relations (rather than merely actions) of
these new objects between themselves and elements of $\O$.  We call
this new algebra the ``universal Cartan calculus'' associated with
$\A$.

However, just as the universal differential calculus approach does not
assume any explicit commutation relations between elements of the
algebra $\A$, and therefore does not coincide with the ``textbook''
differential calculus, neither does our universal Cartan calculus
reduce to the ``textbook'' version, since we take {\em only} the basic
properties of a Hopf algebra as given.  Additional structure on $\O$
(\eg commutation relations) may be incorporated into our formulation
\cite{ixt}; in Section 3 we review the conditions under which this is
possible, and elaborate on how it is accomplished.

We begin by presenting reviews of the basics of the universal
differential calculus and Hopf algebras.

\subsection{Universal Differential Calculus}

(See \cite{Connes,Coq} for a more detailed discussion of the material
in this subsection.)

Let $\A$ be a unital associative algebra over a field $k$, and $\G$
an $\A$-bimodule such that there exists a linear map $\dg :\A
\rightarrow \G$ which satisfies the following:
\begin{eqnarray}
\dg (\IA)&=&0,\nonumber \\
\dg (ab)&=&\dg (a) b + a \dg (b),\label{gamma}
\end{eqnarray}
where $\IA$ is the unit in $\A$, and $a,b\in \A$.  Note that the
latter of these conditions implies that $\G$ is the span of elements
of the form $a\dg (b)$.

As a concrete example, we take $\G$ to be equal to $\ker m$ as a
vector space ($m:\A \otimes \A \rightarrow \A$ is the algebra
multiplication, which we will usually suppress), \ie the span of
elements of the form $\sum_i a_i\otimes b_i$ where $\sum_i a_i b_i
=0$.  $\G$ is made into an $\A$-bimodule by defining the left and
right actions of $\A$ to be $c (\sum_i a_i \otimes b_i)=\sum_i (ca_i
)\otimes b_i$ and $(\sum_i a_i \otimes b_i)c=\sum_i a_i \otimes (b_i
c)$, $c \in \A$.  The $\dg$ which satisfies all the needed conditions
is given by $\dg (a) := \IA \otimes a-a \otimes \IA$.

We now introduce $\O$, the {\em differential envelope associated
with} $\A$; it is the algebra which is spanned by elements of $\A$,
together with formal products of elements of $\G$ modulo the
relations (\ref{gamma}), namely, elements of the form $a_0 \dg (a_1)
\dg (a_2) \ldots \dg (a_p)$.  Such elements are called $p$-forms (\eg
0-forms are elements of $\A$, 1-forms elements of $\G$, {\it
etc.}\footnote{To be precise, we should actually say that \eg 0-forms
are elements of $\iota (\A)$, where $\iota : \A\rightarrow \O$ is the
inclusion map, but we will be glib and suppress this notation
throughout this paper.}).  $\O$ is easily seen to be associative and
unital (with unit $1=\IA$); furthermore, $\dg$ can be extended to a
linear map $\dg :\O \rightarrow \O$ by requiring
\begin{eqnarray}
\dg (1)&=&0,\nonumber \\
\dg^2 (\alpha)&=&0,\nonumber \\
\dg (\alpha \beta)&=&\dg (\alpha) \beta +(-1)^p \alpha \dg
(\beta),\label{Leibniz}
\end{eqnarray}
where $\al, \beta \in \O$, $\al$ a $p$-form.  Thus, $\dg$ maps
$p$-forms to $(p+1)$-forms.  $\dg$ is the {\em exterior derivative} on
$\O$, and we call $(\O,\dg)$ the {\em universal differential
calculus (UDC) associated with} $\A$.

\noindent {\bf Note:}  throughout the remainder of the paper, we shall
use the terms ``0-form'' and ``function'' interchangably to refer to
any element of $\A$.

\subsection{Hopf Algebras}

(See \cite{Maj1}-\cite{Df} for more information about Hopf algebras.)

A Hopf algebra $\A$ is an associative unital algebra (with
multiplication $m$) over a field $k$, equipped with a coproduct
$\Delta:\A \rightarrow \A \otimes \A$, an antipode $S:\A
\rightarrow \A$ (in this paper, we assume the inverse $S^{-1}$ also
exists, although this is not necessarily true for an arbitrary Hopf
algebra), and a counit $\epsilon:\A \rightarrow k$; these maps satisfy
the usual consistency conditions:
\begin{eqnarray}
(\Delta \otimes \id)\Delta(a)&=& (\id \otimes \Delta)\Delta(a),
\nonumber \\
(\epsilon \otimes \id)\Delta(a) = (\id \otimes \epsilon)\Delta(a)
&=&a,\nonumber \\
m(S \otimes \id)\Delta(a) = m(\id \otimes S)\Delta(a)&=&\IA
\epsilon(a),\nonumber \\
\Delta(a b) =\Delta(a) \Delta(b),&&\epsilon(a b) = \epsilon(a)
\epsilon(b),\nonumber \\
\Delta(\IA) = \IA \otimes \IA,&&\epsilon(\IA) = 1_{k},
\end{eqnarray}
for all $a,b \in \A$.  (We will often use Sweedler's \cite{Sw} notation
for the coproduct:
\begin{eqnarray}
\Delta(a) &\equiv& a_{(1)} \otimes a_{(2)},\nonumber\\
(\Delta \otimes \id)\Delta(a) &\equiv& a_{(1)}\otimes a_{(2)}\otimes
a_{(3)},\label{sweedler}
\end{eqnarray}
and so forth, where summation is understood.)

We call two Hopf algebras $\U$ and $\A$ {\em dually paired} if there
exists a nondegenerate inner product $\inprod{\,}{\,}: \U \otimes \A
\rightarrow k$ such that
\begin{eqnarray}
\inprod{xy}{a} & = &\inprod{x \otimes y}{\Delta(a)} \equiv
\inprod{x}{a_{(1)}}\inprod{y}{a_{(2)}}, \label{multinduced}\\
\inprod{x}{a b} & = & \inprod{\Delta(x)}{a \otimes b} \equiv
\inprod{x_{(1)}}{a}\inprod{x_{(2)}}{b}, \label{inducedmult}\\
\inprod{S(x)}{a} & = & \inprod{x}{S(a)},\\
\inprod{x}{\IA}=\epsilon(x),&&\inprod{\IU}{a}=\epsilon(a),
\end{eqnarray}
for all $x,y \in \U$ and $a,b \in \A$.

\section{Universal Cartan Calculus}

The purpose of this section is to generalize the ``classical'' case,
namely the familiar situation where $\O$ is graded-commutative and
the Cartan calculus contains Lie derivatives and inner derivations
which act on $\O$.  Our ``deformed'' version presented here assumes
not only (possible) noncommutativity of $\O$, but also a Hopf
algebraic structure on $\A$.  However, just as in the classical case,
we need specify only how the derivations act on and commute with 0-
and 1-forms; the extension to arbitrary $p$-forms in $\O$ follows
immediately.

We begin with two dually paired Hopf algebras $\A$ and $\U$, and take
$(\O,\dg)$ be the UDC associated with $\A$.  Recall that $\U$ can be
interpreted as an algebra of left-invariant generalized derivations
which act on elements of $\A$ via the action
\begin{equation}
x\trg a = a_{(1)} \inprod{x}{a_{(2)}},
\end{equation}
where $x \in \U$ and $a \in \A$.  The action of $x$ on a product of
functions $a,b\in \A$ is given in terms of the coproduct of $x$:
\begin{equation}
x\trg (ab)  = (x_{(1)} \trg a) (x_{(2)} \trg b).\label{XAB}
\end{equation}
This motivates the introduction of a product structure on the ``cross
product'' algebra $\A\rtimes\U$ \cite{SW}-\cite{Maj2} via the
commutation relation
\begin{equation}
x a = a_{(1)} \inprod{x_{(1)}}{a_{(2)} } x_{(2)}.
\end{equation}

We now introduce for each $x\in \U$ a new object, the {\em Lie
derivative} $\lie_x$; it is linear in $x$, and is a linear map taking
$\O$ into itself such that $p$-forms map to $p$-forms.  Furthermore,
we require that
\begin{equation}
\lie_x \dg = \dg \lie_x.\label{Lie-der}
\end{equation}
This relation allows us to uniquely recover the action of $\lie_x$ on
all of $\O$ from its action on $\A$, \ie 0-forms.  Just as in the
classical case, the action of the Lie derivative on $a\in \A$ is
defined to be the same as that of the corresponding differential
operator, \ie
\begin{equation}
\lie_x(a) =x \trg a=a_{(1)}\inprod{x}{a_{(2)}},\\
\end{equation}
and likewise for its commutation relations with 0-forms:
\begin{equation}
\lie_x a =a_{(1)}\inprod{x_{(1)}}{a_{(2)} }\lie_{x_{(2)}}
=\lie_{x_{(1)}}(a)\lie_{x_{(2)}}.\label{XA}
\end{equation}
{}From (\ref{Lie-der}) and (\ref{XA}) we can find the action on and
commutation relation with a 1-form\footnote{We use parentheses to
delimit operations like $\dg$, $\I_x$ and $\lie_x$, \eg $\dg a =\dg(a)
+ a \dg$.  However, if the limit of the operation is clear from the
context, we will suppress the parentheses, \eg $\dg(\lie_x \dg a)
\equiv\dg(\lie_x(\dg(a)))$.}:
\begin{eqnarray}
\lie_x (\dg a)&=&\dg (a_{(1)})\inprod{x}{a_{(2)}}\nonumber \\
\lie_x \dg(a)&=&\dg(a_{(1)}) \inprod{x_{(1)}}{a_{(2)} } \lie_{x_{(2)}}
=\lie_{x_{(1)}}(\dg a)\lie_{x_{(2)}}.\label{XDA}
\end{eqnarray}

At this point we introduce for each $x\in \U$ the corresponding {\em
inner derivation} $\I_x$.  The guideline for this generalization of
the classical case will be the Cartan identity\footnote{The idea is to
use this identity as long as it is consistent and modify it when
needed.}
\begin{equation}
\lie_x = \I_x  \dg + \dg  \I_x \label{Cartan}
\end{equation}
(so $\I_x$ is linear in $x$).  To find the action of $\I_x$ on $\O$
we can now attempt to use (\ref{Cartan}) in the identity
$\lie_x(a)=\I_x(\dg a) + \dg(\I_x a)$.  We take as an assumption that
the action of $\I_x$ on 0-forms like $a$ vanishes; therefore, we
obtain
\begin{equation}
\I_x(\dg a)= a_{(1)} \inprod{x}{a_{(2)} }.
\label{incomplete}
\end{equation}
However, this cannot be true for any $x \in \U$ because, by
assumption, $\dg (1)=0$.  From (\ref{incomplete}), $\I_x (\dg
1)=1\epsilon (x)$, which is not necessarily zero.  We see that the
trouble arises when dealing with $x \in \U$ with $\epsilon(x) \neq 0$.
Noting that $\epsilon(x -\IU \epsilon(x)) = 0$, we modify equation
(\ref{incomplete}) to read
\begin{equation}
\I_x(\dg a) = a_{(1)} \inprod{x-\IU \epsilon(x)}{a_{(2)}},
\label{ida}
\end{equation}
so that $\I_x(\dg 1)$ does indeed vanish for all $x$.  Also note that
this requires the consistency condition
\begin{equation}
\I_{\IU} \equiv  0.
\end{equation}
To allow for all $x \in \U$ with nonzero counit, we also need to
modify equation (\ref{Cartan}) to
\begin{equation}
\lie_{x - \IU \epsilon(x)} = \I_x  \dg + \dg  \I_x,
\end{equation}
or, in view of (\ref{XA}), identifying $\lie_{\IU} \equiv \bid$ and
using the linearity of the Lie derivative,
\begin{equation}
\lie_x = \I_x \dg + \epsilon(x)\bid + \dg \I_x \label{uCi}
\end{equation}
(here $\bid$ is the identity map on $\O$, and therefore the unit in
the algebra of generalized derivations).  We call this the {\em
universal Cartan identity}.

To find the complete commutation relations of $\I_x$ with elements of
$\O$ rather than just its action on them, we need only find out how
$\I_x$ moves through 0- and 1-forms.  Both of these can be found by
commuting $\lie_x$ through a function $a\in \A$, using (\ref{XA}) and
(\ref{uCi}): the left-hand side of the former equation gives (using
the Leibniz rule)
\begin{equation}
\lie_x a =\I_x \dg (a)+\I_x a \dg +\epsilon (x)a +\dg \I_x a
\end{equation}
and the right-hand side gives
\begin{eqnarray}
\lefteqn{a_{(1)}\inprod{x_{(1)}}{a_{(2)}}\lie_{x_{(2)}} =} \nonumber \\
& &a_{(1)}\inprod{x_{(1)}}{a_{(2)}}\dg \I_{x_{(2)}} \\
& &+a_{(1)}\inprod{x}{a_{(2)}}+a_{(1)}\inprod{x_{(1)}}{a_{(2)}}\I_{x_{(2)}}
\dg.\nonumber
\end{eqnarray}
Equating the two and using (\ref{Leibniz}), (\ref{Lie-der}),
(\ref{ida}) and $\I_x (a)=0$, we obtain
\begin{equation}
\I_x \dg (a)-\I_x (\dg a)+\lie_{x_{(1)}} (\dg a)\I_{x_{(2)}}=
\acomm{-\I_x a+\I_x (a)+\lie_{x_{(1)}} (a)\I_{x_{(2)}}}{\dg}.\label{anti}
\end{equation}
Therefore, we propose the commutation relation
\begin{equation}
\I_x \al = \I_x (\al)+(-1)^p \lie_{x_{(1)}}(\al)\I_{x_{(2)}}\label{IXA}
\end{equation}
for any 0- or 1-form $\al$, so that both sides of (\ref{anti}) vanish.

Missing in our list are commutation relations of Lie derivatives with
themselves and inner derivations.  To find the $\lie$-$\lie$ relations,
we note that it follows from the Hopf algebra axioms that the product
in $\U$ can be expressed as
\begin{equation}
xy \equiv (x_{(1)}\ad y)x_{(2)},
\end{equation}
where $x\ad y:=x_{(1)}yS(x_{(2)})$ is the adjoint action on $\U$.  As
before, we extend the properties of the elements of $\U$ to those of
the corresponding Lie derivatives to find
\begin{equation}
\lie_x \lie_y =\lie_{(x_{(1)}\tad y)}\lie_{x_{(2)}},
\end{equation}
and therefore, using (\ref{uCi}),
\begin{equation}
\lie_x \I_y  =\I_{(x_{(1)}\tad y)}\lie_{x_{(2)}}.\label{Lie-in}
\end{equation}
(It would seem that (\ref{uCi}) could also give the relation
$$
\I_x \lie_y = \lie_{(x_{(1)}\tad y)}\I_{x_{(2)}}+\I_{(x-\IU
\epsilon(x)) \tad y},
$$
but this is inconsistent with the commutation relation (\ref{IXA}).)

To recap our results of this section, we present a summary of the
actions of the Lie derivatives and inner derivations with 0- and
1-forms:
\begin{eqnarray}
\lie_x (a) & = & a_{(1)} \inprod{x}{a_{(2)}},\\
\lie_x (\dg a) & = & \dg(a_{(1)}) \inprod{x}{a_{(2)}},\\
\I_x (a) & = &0,\\
\I_x (\dg a) & = & a_{(1)} \inprod{x -\IU \epsilon(x)}{a_{(2)}},
\end{eqnarray}
where, as usual, $x\in \U$, $a\in \A$.  The commutation relations are
therefore
\begin{eqnarray}
\lie_x \al &=& \lie_{x_{(1)}}(\al )\lie_{x_{(2)}},\\
\I_x \al &=& \I_x (\al )+(-1)^p \lie_{x_{(1)}}(\al )\I_{x_{(2)}},
\end{eqnarray}
where $\al \in \O$ is a $p$-form.  (The actions and commutation
relations for $\dg$ were already given when the UDC was introduced.)
Finally, here are the relations between the derivations themselves:
\begin{eqnarray}
\acomm{\dg}{\dg}&=&0,\\
\comm{\dg}{\lie_x}&=&0,\\
\acomm{\dg}{\I_x}&=&\lie_x -\epsilon(x)\bid,\\
\lie_x \lie_y &=& \lie_{(x_{(1)}\tad y)}\lie_{x_{(2)}}\\
\lie_x \I_y &=& \I_{(x_{(1)}\tad y)}\lie_{x_{(2)}}
\end{eqnarray}

Note that at this point we do not have $\I$-$\I$ commutation
relations.  This is not a problem; an expression like $\I_x \I_y$ is
simply an element of the calculus whose action on and commutation
relations with $p$-forms are perfectly well-defined.  This is much
like the fact that $\dg(a)\dg(b)$ and $\dg(b)\dg(a)$ are simply
elements of $\O$; the UDC does not a priori impose relations such as
$\dg(a)\dg(b)+\dg(b)\dg(a)\equiv 0$ (unlike the ``classical'' case).
However, we will see in Section 3 that such restrictions between
elements of $\O$ are possible in some cases, and we will comment on
the possibility of $\I$-$\I$ comutation relations.

\subsection{Cartan-Maurer Forms}

The most general left-invariant 1-form can be written \cite{Wor}
\begin{equation}
\om_b := S(b_{(1)}) \dg(b_{(2)}) = - \dg(S b_{(1)} ) b_{(2)},
\end{equation}
corresponding to a function $b \in \A$. (To connect with the classical
case, if $\A$ is an $m \times m$ matrix representation of some Lie
group with $\Delta (g^i{}_j)=g^i{}_k \otimes g^k{}_j$, $S(g^i{}_j)=
(g^{-1})^i{}_j$ and $\epsilon (g^i{}_j)=\delta^i{}_j$ for $g\in \A$,
then $\om_g = g^{-1} \dg(g)$, \ie $\om_g$ is the well-known
left-invariant Cartan-Maurer form.)  Here is a nice formula for the
exterior derivative of $\om_b$:
\begin{eqnarray}
\dg(\om_b) & = & \dg(S b_{(1)} ) \dg(b_{(2)})\nonumber \\
&=& \dg(S b_{(1)} ) b_{(2)} S(b_{(3)}) \dg(b_{(4)})\nonumber \\
&=& - \om_{b_{(1)}} \om_{b_{(2)}}.\label{om-om}
\end{eqnarray}
The Lie derivative on $\om_b$ is
\begin{eqnarray}
\lie_x(\om_b)&=&\lie_{x_{(1)}}(S b_{(1)} ) \lie_{x_{(2)}}(\dg b_{(2)})
\nonumber\\
&=&\inprod{x_{(1)}}{S(b_{(1)})} S(b_{(2)}) \dg(b_{(3)})
\inprod{x_{(2)}}{b_{(4)}}\nonumber \\
&=&\om_{b_{(2)}} \inprod{x}{S(b_{(1)})b_{(3)}}. \label{XOM}
\end{eqnarray}
The contraction of left-invariant forms with $\I_x$ gives a number in
the field $k$, rather than a function in $\A$ (as was the case for
$\dg(a)$):
\begin{eqnarray}
\I_x(\om_b) & = & \I_x(- \dg(S b_{(1)} ) b_{(2)})\nonumber \\
&=& - \I_x(\dg S b_{(1)} ) b_{(2)}\nonumber \\
& = & -\inprod{x -\IU \epsilon(x)}{S(b_{(1)})} S(b_{(2)}) b_{(3)}
\nonumber \\
&=& -\inprod{x}{S(b)} + \epsilon(x) \epsilon(b).\label{IOM}
\end{eqnarray}
(This result is a consequence of the fact that $\U$ was interpreted as
an algebra of left-invariant differential operators, so $\I_x (\om_b)$
must be a left-invariant 0-form, \ie proportional to $1$.)

As an exercise, as well as a demonstration of the consistency of our
results, we will compute the same expression in a different way:
\begin{eqnarray}
\I_x(\om_b) & = & \I_x(S(b_{(1)}) \dg(b_{(2)}))\nonumber \\
& =& \inprod{x_{(1)}}{S(b_{(1)})} S(b_{(2)}) \I_{x_{(2)}}(\dg b_{(2)})
\nonumber \\
& = & \inprod{x_{(1)}}{S(b_{(1)})} S(b_{(2)}) b_{(3)}
\inprod{x_{(2)} - \IU \epsilon(x_{(2)})}{b_{(4)}}\nonumber \\
& =& \inprod{x_{(1)}}{S(b_{(1)})}\inprod{x_{(2)} - \IU
\epsilon(x_{(2)})}{b_{(2)}} \nonumber \\
&=&  \epsilon(x) \epsilon(b) -\inprod{x}{S(b)}.
\end{eqnarray}

As a final observation, if $\{e_i\}$ and $\{f^i\}$ are, respectively,
(countable) bases of $\U$ and $\A$ with $\inprod{e_i}{f^j}=
\delta_i^j$, the action of $\dg$ on functions $a \in \A$ may be
expressed as
\begin{equation}
\dg (a) =\lie_{e_i}(a)\om_{f^i}=-\om_{S^{-1}(f^i)} \lie_{e_i}(a);
\end{equation}
so that the Cartan-Maurer forms form a left-invariant basis for $\G$.

\section{Further Commutation Relations}

\subsection{Generalized Differential Calculus}

Recall that, so far, the only commutation relations we have in $\O$
are those which follow from the Leibniz rule (\ref{Leibniz}); we
assume nothing else.  Here we review the standard method of
introducing nontrivial commutation relations into the differential
envelope which maintains the covariance properties we have chosen (\eg
left-invariance of the Cartan-Maurer forms) \cite{Wor}.

Let $\K\equiv\ker\epsilon\subset\A$, and suppose there exists a
subalgebra $\R \subset \A$ which satisfies
\begin{enumerate}
\item $\R\subseteq\K$,
\item $\R \A \subseteq \R$,
\item $\Ad (\R) \subseteq \R \otimes \A$
\end{enumerate}
(where $\Ad (a)=a_{(2)}\otimes S(a_{(1)})a_{(3)}$ is the adjoint
coaction on $\A$).  We define the submodule $\N\subseteq \G$ as the
space spanned by 1-forms of the form $a\om_r$, where $a\in \A$ and
$r\in\R$.  The above properties of $\R$ imply properties of $\N$: (1)
and (2) give $\N\A\subseteq\N$, and (3) gives $\DA (\N)\subseteq \N
\otimes \A$.  Such an $\R$ always exists; $\{ 0\}$ and $\K$ both
satisfy all three conditions.

With $\R$ as above, we can construct the $\A$-module $\GN := \G / \N$.
When $\R=\{ 0\}$, and therefore $\N=\{ 0\}$, the only commutation
relations between elements of $\A$ and $\GN$ are those allowed by the
Leibniz rule, and we recover the UDC; when $\R=\K$, $\N=\G$, so
$\GN=\{ 0\}$, and we end up with a trivial differential calculus.
However, if there exists an $\R$ in between these two extreme cases,
then there are additional commutation relations between elements of
$\GN$, namely those given by $\om_r\simeq 0$ for $r\in\R$ ($\simeq$
being the equivalence relation in $\GN$).  Furthermore, we find
explicit commutation relations between elements of $\GN$ by using
(\ref{om-om}) and the properties of $\R$, \ie
$\om_{r_{(1)}}\om_{r_{(2)}}\simeq 0$.  Therefore, we no longer have a
UDC, but rather a differential envelope with nontrivial commutation
relations which is constructed using $\A$ and $\GN$; we refer to this
envelope as $\ON$, and the pair $(\ON,\dg)$ is referred to as the {\em
generalized differential calculus (GDC) associated with} $\A$ {\em
and} $\R$.

\subsection{Generalized Cartan Calculus}

How do we incorporate our Cartan calculus into this scheme?  We start
by defining a subspace $\T\subset\U$, given by
\begin{equation}
\T:=\{ x\in \U |\epsilon(x)=0;\, \inprod{x}{S(r)}=0,\, r\in\R\}.
\end{equation}
It is easily seen that the defining properties for $\R$ imply,
respectively,
\begin{enumerate}
\item $\IU \not\in \T$,
\item $\Delta (\T)\subseteq\U\otimes (\T\oplus\IU)$,
\item $\U \ad \T\subseteq \T$.
\end{enumerate}
Note that for $x\in\T$ and $a\in\A$,
\begin{equation}
\I_x (\om_a )=-\inprod{x}{S(a)}.\label{dual}
\end{equation}
Suppose this vanishes; then either $x=0$, $a=\IA$, or $a\in\R$.
Therefore, if we restrict $a$ to be in $\KR$, then the vanishing of
(\ref{dual}) implies that $x=0$ or $a=0$, \ie the inner product
$\dinprod{\,}{\,}:\T\otimes\KR \rightarrow k$ defined by
\begin{equation}
\dinprod{x}{a}:=-\inprod{x}{S(a)}\label{dinprod}
\end{equation}
is nondegenerate.  Hence, $\T$ and $\KR$ are dual to one another.  The
nondegeneracy of (\ref{dinprod}) guarantees that the map from
$\KR\rightarrow\T^{*}$ given by $a\mapsto\om_a$ is bijective, insuring
that $\GN$ is the space of all 1-forms over $\A$.  Therefore, to
consistently define our Cartan calculus on all of $\ON$, we must
restrict the arguments of the Lie derivative and inner derivation from
$\U$ to $\T$, and the argument of $\om$ from $\A$ to $\KR$.

(As an example of how this works, note that for $x\in\T$ and $a\om_r
\in\N$,
\begin{eqnarray}
\lie_x a\om_r &=& a_{(1)}\om_{r_{(2)}}\inprod{x_{(1)}}{a_{(2)}
S(r_{(1)})r_{(3)}}\lie_{x_{(2)}},\nonumber \\
\I_x a\om_r &=&-a_{(1)}\om_{r_{(2)}}\inprod{x_{(1)}}{a_{(2)}
S(r_{(1)})r_{(3)}}\I_{x_{(2)}}.
\end{eqnarray}
Property (3) of $\R$ guarantees that $a_{(1)}\om_{r_{(2)}}
\inprod{x}{a_{(2)}S(r_{(1)})r_{(3)}}\in \N$ for all $x\in\U$, so both
sides of the two preceding equations are $\simeq 0$ in $\GN$.)

A specific example where a nontrivial $\R$ exists is the case of the
quantum group $GL_q (N)$.  It was shown in \cite{Schirr} that
nontrivial commutation relations between 0- and 1-forms can be
expressed using the $GL_q (N)$ quantum matrix $A$ and R-matrix $R$;
these are obtained using the above prescription, with $\R$ being the
subalgebra of $GL_q (N)$ generated by the $N^4$ elements
\begin{equation}
\matrix{r}{ij}{k\ell}:=\matrix{(A_1 A_2 -A_2 -R^{-1}A_1 R_{21}^{-1}
+R^{-1}R_{21}^{-1})}{ij}{k\ell}.
\end{equation}
The consistency of the resultant relation is entirely dependent upon
the fact that the R-matrix for $GL_q (N)$ satisfies a quadratic
characteristic equation.

When the Cartan calculus for this case was found in \cite{SWZ},
consistent $\I$-$\I$ commutation relations, written in terms of the
R-matrix, were given.  These relations were also dependent upon the
form of the $GL_q (N)$ R-matrix characteristic equation.  However, we
have not yet found a method for explicitly expressing such $\I$-$\I$
relations in a form depending manifestly on $\R$, \ie in the flavor of
$\om_{r_{(1)}}\om_{r_{(2)}}\simeq 0$.

\section*{Acknowledgement}

The authors would like to express their utmost gratitude to Prof.
Bruno Zumino for many useful discussions, as well as for his support
of and contributions to this work.

This work was supported in part by the Director, Office of Energy
Research, Office of High Energy and Nuclear Physics, Division of High
Energy Physics of the U.S. Department of Energy under Contract
DE-AC03-76SF00098 and in part by the National Science Foundation under
grants PHY-90-21139 and PHY-89-04035.


\begin{thebibliography}{199}
\bibitem{Wor} S. L. Woronowicz, {\it Commun. Math. Phys.} {\bf 122}
125 (1989)
\bibitem{RTF} N. Yu. Reshetikhin, L. A. Takhtadzhyan and L. D.
Faddeev, {\it Leningrad Math. J.} {\bf 1} 193 (1990)
\bibitem{SWZ} P. Schupp, P. Watts and B. Zumino, {\it Lett. Math.
Phys.} {\bf 25} 139 (1992)
\bibitem{B} D. Bernard, {\it Prog. Theor. Phys. Supp.} {\bf 102} 49
(1990)
\bibitem{CC} P. Aschieri and L. Castellani, {\it Int. J. Mod. Phys.}
{\bf A8} 1667 (1993)
\bibitem{CW} U. Carow-Watamura, M. Schlieker, S. Watamura and W.
Weich, {\it Commun. Math. Phys.} {\bf 142} 605 (1991)
\bibitem{Connes} A. Connes, {\it Pub. Math. (IHES)} {\bf 62} 257 (1985)
\bibitem{ixt} P. Schupp, P. Watts and B. Zumino, preprint LBL-34833,
UCB-PTH-93/32, hep-th 9312073 (1993), to appear (in short form) in
{\it Proc.} XXII$^{th}$ {\it Int. Conf. Diff. Geom. Meth. Theor.
Phys., Ixtapa, Mexico, Sept. 1993}
\bibitem{Coq} R. Coquereaux, preprint CPT-88/P-2147 (1988)
\bibitem{Maj1} S. Majid, {\it Int. J. Mod. Phys.} {\bf A5} 1 (1990)
\bibitem{A} E. Abe, \underline{Hopf Algebras}, Cambridge Univ. Press,
1977
\bibitem{Sw} M. E. Sweedler, \underline{Hopf Algebras}, Benjamin, 1969
\bibitem{Df} V. G. Drinfel'd, {\it Proc. Int. Congr. Math., Berkeley,
1985} 798 (1986)
\bibitem{SW} C. Chryssomalakos, P. Schupp and P. Watts, preprint
LBL-33274, UCB-PTH-92/42, hep-th 9310100 (1993)
\bibitem{SWZ3} P. Schupp, P. Watts and B. Zumino, {\it Commun. Math.
Phys.} {\bf 157} 305 (1993)
\bibitem{Maj2} S. Majid, {\it Int. J. Mod. Phys.} {\bf A8} 4521 (1993)
\bibitem{Schirr} A. Schirrmacher, preprint MPI-PTH-91-117 (1991),
presented at 1$^{st}$ Max Born Symp. Theor. Phys., Wroclaw, Poland,
Sept. 1991
\end{thebibliography}
\end{document}